\documentclass[12pt]{article}
\usepackage{amssymb}
\usepackage{graphicx}
\usepackage{epsfig}
\usepackage{wrapfig}

\newcommand{\tr}{\mbox{tr}}

\newcommand{\lapproxeq}{\lower .7ex\hbox{$\;\stackrel{\textstyle
<}{\sim}\;$}}
\newcommand{\gapproxeq}{\lower .7ex\hbox{$\;\stackrel{\textstyle
>}{\sim}\;$}}
\newcommand{\stackdown}[2]{\lower 1.4ex\hbox{$\;\stackrel{\textstyle{#1}}
{\scriptstyle{#2}}\;$}}
\newcommand{\beq}{\begin{equation}}
\newcommand{\eeq}{\end{equation}}
\newcommand{\bea}{\begin{eqnarray}}
\newcommand{\eea}{\end{eqnarray}}

\newcommand{\real}{{\bb R}} 

\def\(#1{ ^{(#1)} }

\font\mybb=msbm10 at 12pt
\def\bb#1{\hbox{\mybb#1}}

\def\beq{\begin{equation}}
\def\eeq{\end{equation}}
\def\e{{\rm e}}

\makeatletter
\def\slash{\@ifnextchar[{\fmsl@sh}{\fmsl@sh[0mu]}}
\def\fmsl@sh[#1]#2{%
  \mathchoice
    {\@fmsl@sh\displaystyle{#1}{#2}}%
    {\@fmsl@sh\textstyle{#1}{#2}}%
    {\@fmsl@sh\scriptstyle{#1}{#2}}%
    {\@fmsl@sh\scriptscriptstyle{#1}{#2}}}
\def\@fmsl@sh#1#2#3{\m@th\ooalign{$\hfil#1\mkern#2/\hfil$\crcr$#1#3$}}
\makeatother

\newcommand{\nn}{\nonumber}
\newcommand{\eqn}[1]{(\ref{#1})}

\def\beq{\begin{equation}}
\def\eeq{\end{equation}}

\catcode`\@=11 

\def\lsim{\mathrel{\mathpalette\@versim<}}
\def\gsim{\mathrel{\mathpalette\@versim>}}
\def\@versim#1#2{\vcenter{\offinterlineskip
    \ialign{$\m@th#1\hfil##\hfil$\crcr#2\crcr\sim\crcr } }}

\def\t1{{\tilde 1}}

\def\slash#1{#1\hskip-6pt/\hskip6pt}

\def\to{\rightarrow}

\begin{document}

\begin{titlepage}

\begin{flushright}
gr-qc/0312044\\
CERN-TH/2003-296\\
MIFP-03-26\\
\end{flushright}
\vspace{0.4cm}

\begin{centering}

{\Large {\bf Space-Time Foam may Violate the Principle of Equivalence}}
\vspace{0.4cm}

{\bf John Ellis$^a$}, {\bf N.E.~Mavromatos$^{b,c}$},
{\bf D.V.~Nanopoulos$^d$} and
{\bf A.S.~Sakharov$^{a,e,f}$}

\vspace{0.4cm}
$^a$ {\it CERN, Theory Division, CH-1211 Geneva 23, Switzerland.}

$^b$ {\it King's College London, University of London, Department of Physics,
Strand WC2R 2LS, London, U.K.}

$^c$ {\it Departamento de F\'isica T\'eorica, Universidad de Valencia,
E-46100, Burjassot, Valencia, Spain.}

$^d$ {\it George P. and Cynthia W. Mitchell Institute for Fundamental
Physics, Texas A\&M University, College Station, TX 77843, USA; \\

Astroparticle Physics Group, Houston
Advanced Research Center (HARC),
Mitchell Campus,
Woodlands, TX~77381, USA; \\
Academy of Athens,
Academy of Athens,
Division of Natural Sciences, 28~Panepistimiou Avenue, Athens 10679,
Greece.}

$^e$  {\it Swiss Institute of Technology, ETH-Z\"urich, 8093 Z\"urich,
Switzerland.}

$^f$ {\it INFN Laboratory Nazionali del Gran Sasso, SS. 17bis
67010 Assergi (L'Aquila), Italy.}
\vspace{0.4cm}

{\bf Abstract}

\end{centering}

The interactions of different particle species with the foamy space-time
fluctuations expected in quantum gravity theories may not be universal, in
which case different types of energetic particles may violate Lorentz
invariance by varying amounts, violating the equivalence principle. We
illustrate this possibility in two different models of space-time foam
based on D-particle fluctuations in either flat Minkowski space or a stack
of intersecting D-branes. Both models suggest that Lorentz invariance
could be violated for energetic particles that do not carry conserved
charges, such as photons, whereas charged particles such electrons would
propagate in a Lorentz-inavariant way. The D-brane model further suggests
that gluon propagation might violate Lorentz invariance, but not
neutrinos. We argue that these conclusions hold at both the tree
(lowest-genus) and loop (higher-genus) levels, and discuss their
implications for the phenomenology of quantum gravity.

\end{titlepage}

\section{Introduction}

The reconciliation of general relativity and quantum field theory in a
true quantum theory of gravity remains elusive. Perhaps one or the other,
or both, of these fundamental theories undergoes some modification when
embedded in quantum gravity? A possible hint for a need to modify quantum
field theory may have been provided by black-hole
thermodynamics~\cite{Bek}. In general, information is lost across a
horizon, such as that surrounding a black hole, which must therefore be
described as a mixed quantum-mechanical state. However, one can imagine
preparing a black hole from a pure quantum-mechanical initial state, so
must we formulate~\cite{impure} a theory that, unlike conventional quantum
mechanics and quantum field theory, allows pure states to evolve into
mixed states? Another suggestion is that Lorentz symmetry may be broken in
a quantum theory of gravity.  In such a theory, the vacuum presumably
needs to be treated as a dynamical medium - a space-time foam containing
many evanescent fluctuations. Energetic particles may induce recoil
effects on structures in the foam, and these may, in return, induce
back-reaction effects on the the particles' propagation. Do these cause
energetic particles to travel at less than the conventional speed of
light?

In this paper we propose a third possible signature of quantum gravity,
namely a violation of the equivalence principle~\cite{emns}, in the sense
that different types of ultra-relativistic particles may propagate in
different ways. Within the general quantum-gravitational framework
outlined above, such equivalence breaking might arise if the back-reaction
of the recoiling space-time foam is non-universal for different particle
species.  This would be unexpected if the structures that appear and
disappear in the space-time foam have purely gravitational interactions
with the propagating matter particles. However, there might also be
non-gravitational interactions with structures in the space-time foam,
which need not be universal. In this case the back-reactions might depend
on the species of the propagating particle~\cite{ems}.

We demonstrate that just such a feature appears in a D-brane model of
space-time foam~\cite{emn} that we have used previously to motivate the
suggestion that energetic photons might travel at {\it less} than the
standard velocity of light measured for low-energy
photons~\cite{aemn,nature}. We show that the same mechanism that {\it
slows down} energetic photons, producing an energy-dependent {\it in
vacuo} refractive index, is {\it inoperative} for relativistic particles
carrying a conserved charge, such as electrons.  This is because this
approach models the structures in the space-time foam as D-particles, and
neither they nor their possible excitations can carry electric charge.

We were led to this observation that the principle of equivalence might be
violated for ultra-relativistic particles by considering the severe
constraints on Lorentz violation for {\it electrons} \cite{jlm,ems} that
are imposed by the apparent observation of synchrotron radiation from the
Crab Nebula~\cite{jlm}.  We observed \cite{ems} that this impressive
constraint does not apply to {\it photons}, and we then realized that our
D-brane evades the Crab Nebula constraint just because it predicts Lorentz
invariance for electrons, {\it in contrast} to photons~\cite{emns}.

In this paper, we focus on the explanation of this equivalence breaking in
two variants of the D-particle model, one based on string propagation in
Minkowski space and the other invoking intersecting stacks of
D-branes~\cite{polchinski}. The latter suggests that the propagation of
energetic gluons might violate Lorentz invariance, but not that of
energetic neutrinos. We argue that these conclusions should hold also at
higher orders in the string genus (loop) expansion. Finally, we comment on
the interpretation of the constraint on quantum gravity provided by
synchrotron radiation from the Crab Nebula, and on other implications for
quantum-gravity phenomenology.

\section{Space-Time Foam and the Violation of the Equivalence Principle}

It may at first sight seem paradoxical that space-time foam could violate
the equivalence principle, since one expects gravity to exhibit universal
properties at low energies. This low-energy universality is a
well-established feature of string theory, but it is not guaranteed that
this universality should extend to high energies. For example, in brane
physics, excitations with non-zero charges, e.g., under the Standard Model
gauge group, are represented by {\it open} strings with their ends
attached with Dirichlet boundary conditions to a brane~\cite{polchinski}.
Only {\it neutral, closed-string} excitations are allowed to propagate in
the bulk space transverse to the brane. This framework is manifestly
non-universal in general.

We display the advertised non-universal violation of the equivalence
principle for ultra-relativistic particles in two different ways in the
following two subsections. The first demonstration is outside the
brane-world framework, and the second is within it.

\begin{figure} [ht]
\begin{center}
\epsfig{file=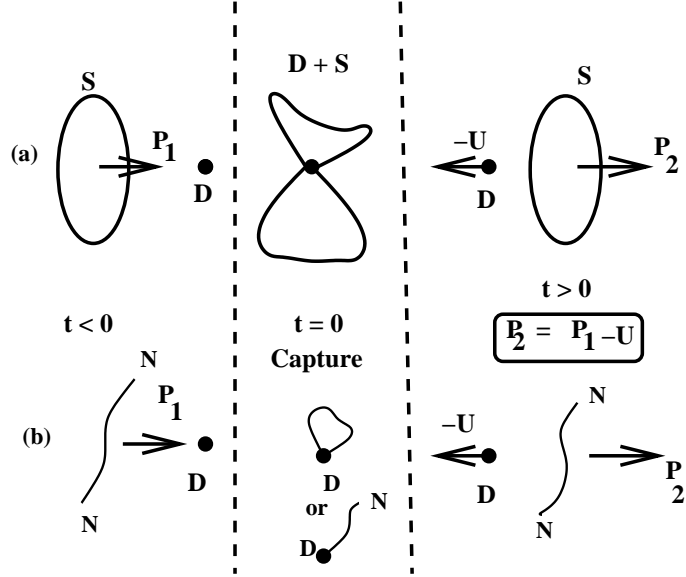,width=0.6\textwidth}
\end{center}
\caption{{\it  
In the Liouville model of space-time foam~\cite{emn}, only a string
particle (S) neutral under the (unbroken) Standard Model group can be
captured by a D-particle defect (D) in space time. This results in a
modified (subluminal) dispersion relation for S, as a consequence of the
recoil of the defect D in a direction opposite to the incident beam. 
Drawing (a) shows the capture of a closed string S by the defect, and 
drawing (b) shows the capture of an open string S by the defect.  
All capture processes involve a change in the boundary conditions of the
respective $\sigma$-model fields on open world sheets.
}}
\label{fig:recoil}
\end{figure}

\subsection{D-Particle Model of Space-Time Foam}

In the original model of ~\cite{emn}, one treats observable 
particles as closed-string states propagating through a `bulk' 
space-time. Space-time 
foam as comprised of point-like defects, each of which is modelled as a
zero-space-dimensional D0-brane or D-particle, embedded in this four- (or 
higher-)
dimensional {\it bulk} space-time, which represents the Minkowski
space-time we observe.  An energetic particle propagating with momentum
$p_i$ may strike a relatively heavy defect with mass $\sim M_s$, the
string scale with a coupling $g_s$, causing it to recoil, as illustrated
in Fig.~\ref{fig:recoil}. This process was discussed in detail
in~\cite{emn}, where it was shown that the associated back-reaction
effects in target space lead to the following energy-dependent modified 
effective metric `felt' by the energetic particle:
\begin{equation}
G_{00}=-1~; \quad G_{ij}=\delta_{ij}~; \quad G_{0i} = g_s {\Delta p_i 
\over M_s} \sim
\frac{1}{2}g_s {p_i \over M_s} \sim {p_i \over {\cal M}},
\label{metricrec}
\end{equation}
where $\Delta p_i$ is the momentum transfer during the scattering 
process and $M_s/g_s = {\cal M}$ is the effective quantum-gravity 
scale~\cite{emn}.

In more detail, in the recoil/back-scattering process illustrated in
Fig.~\ref{fig:recoil}, a closed-string (matter) state - we recall that in
this model there are no open strings in the bulk - is momentarily {\it
captured} by the defect. It then {\it splits} into two open-string
excitations with their ends attached in the defect, which then {\it
recombine} to emit a closed-string state. This will, in general, have a
new momentum, with the defect itself recoiling in a direction {\it
opposite} to that of the incident particle.

The corresponding target-space quantum dynamics was discussed
in~\cite{szabo} from a world-sheet $\sigma$-model point of view, in which
the Galilean-boosted D-particle has position
\beq
Y_i(x^0) = Y_i + U_i\,x^0,
\label{boost}
\eeq
where we have assumed a non-relativistic heavy D-particle with velocity
$U_i$. The mathematical formulation of the capture/recoil process involves
non-trivial impulse operators, as discussed in~\cite{szabo}. One may
incorporate such excitations in the $\sigma$-model path integral by
considering the world-sheet genus expansion of a matrix $\sigma$ model. An
analysis of the annulus amplitude reveals that there are logarithmic
divergences arising from integrations over the modular parameter $q$ of
the form $\int dq/q$~\cite{kogmav,recoil}, describing appropriate
summations of world-sheet annulus diagrams. These divergences can be
removed by replacing the velocity operator in \eqn{boost} by
$\lim_{\epsilon \to 0^+}U_i D(x^0;\epsilon)$, where
\beq
D(x^0;\epsilon)=x^0\,\Theta(x^0;\epsilon)
\label{Depsilonop}
\eeq
and
\beq
\Theta(s;\epsilon)=\frac1{2\pi
i}\int_{-\infty}^\infty\frac{dq}{q-i\epsilon}~\e^{iqs}
\label{Thetafnreg}
\eeq
is the regulated Heaviside step function, with $\Theta(s) \equiv
\lim_{\epsilon \to 0^+} \Theta(s;\epsilon) = 0$ for $s<0$ and $\Theta(s) =
1$ for $s > 0$. The infinitesimal parameter $\epsilon$ regulates the
ambiguous value of $\Theta(s)$ at $s=0$, and the integral representation
\eqn{Thetafnreg} is used because $x^0$ is eventually elevated to a quantum
operator.

When this velocity term is inserted into the boundary integral of the
$\sigma$-model action, the $\epsilon \to 0^+$ divergences arising from the
regulated step function can be used to cancel the logarithmic divergences of
the annulus amplitudes \cite{kogmav,recoil}. This relates the target-space 
regularization parameter $\epsilon$ to the world-sheet ultraviolet scale
$\Lambda$ by~\cite{kmw}
\beq
\epsilon^{-2}=-2\alpha'\log\Lambda
\label{scalerel}
\eeq
This new velocity operator is called the impulse operator~\cite{recoil},
and has non-zero matrix elements between different states of the
D-particle.  It describes recoil effects from the emission or scattering
of closed string states off the D-particle, and, in the impulse
approximation, it ensures that (classically) the D-particle starts moving
only at time $x^0 = 0$. 

But this is not all that is required. The operator
\eqn{Depsilonop} on its own does not lead to a closed conformal algebra.
Computing its operator-product expansion with the stress-energy tensor
shows~\cite{kmw} that it is only the {\it pair} of operators
$D(x^0;\epsilon), C(x^0;\epsilon)$, where
\beq
C(x^0;\epsilon) = \epsilon\,\Theta(x^0;\epsilon)
\label{Cepsilonop}
\eeq
that define an algebra that is closed under the action of the world-sheet
stress-energy tensor. They form a pair of logarithmic operators of the
conformal field theory~\cite{gurarie}. Thus, in order to maintain
conformal invariance of the world-sheet theory, one cannot work just with
the operator \eqn{Depsilonop}, because \eqn{Cepsilonop} will be induced by
conformal transformations. If we rescale the world-sheet cutoff
\beq
\Lambda \to \Lambda' = \Lambda\,\e^{-t/\sqrt{\alpha'}}
\label{Lambdarescale}
\eeq
by a linear renormalization group scale $t$, then \eqn{scalerel} induces a
transformation
\beq
\epsilon^2 
\to \epsilon'^2 = \frac{\epsilon^2}{1 - 4\sqrt{\alpha'}\,\epsilon^2\,t},
\label{epsilonrescale}
\eeq
and we find
\beq
D(x^0;\epsilon')=D(x^0;\epsilon)+t\,\sqrt{\alpha'}\,C(x^0;\epsilon)~~~~~~,~~~~~~
C(x^0;\epsilon')=C(x^0;\epsilon).
\label{DCscaletransfs}
\eeq
If we now modify the initial position of the D-particle to
$\lim_{\epsilon \to 0^+} \sqrt{\alpha'}\,Y_i C(x^0;\epsilon)$, then 
this scale transformation
will induce, by conformal invariance, a transformation of the velocities and
positions as follows:
\beq
U_i\to U_i~~~~~~,~~~~~~Y_i\to Y_i+U_i\,t,
\label{Galileanboost}
\eeq
i.e., a Galilean evolution of the D-particle in target space.

In order to incorporate properly the non-trivial dynamics of the 
D-particle, one must
therefore consider, instead of \eqn{boost}, the recoil operator
\beq
Y_i(x^0) = \lim_{\epsilon\to0^+}\left(\sqrt{\alpha'}\,
Y_i C(x^0;\epsilon) + U_i D(x^0;\epsilon)\right).
\label{recoilop}
\eeq
The conformal algebra reveals that the operators \eqn{Depsilonop} and
\eqn{Cepsilonop} have the same conformal dimension~\cite{kmw}
\beq
\Delta_\epsilon=-\alpha'|\epsilon|^2/2,
\label{Deltaep}
\eeq
which vanishes as $\epsilon \to 0^+$. For finite $\epsilon$, the operator
\eqn{recoilop} yields a deformation of the world-sheet action of conformal
dimension $1-\alpha'|\epsilon|^2/2$. This describes a relevant deformation
of the $\sigma$-model, and the resulting string theory is therefore
non-critical. It follows from \eqn{Galileanboost} that the corresponding
RGE $\beta$ functions are
\beq
\beta_{Y_i}=\Delta_\epsilon
Y_i+\sqrt{\alpha'}\,U_i~~~~~~,~~~~~~\beta_{U_i}=\Delta_\epsilon U_i.
\label{betaepsilon}
\eeq
As the `dressing' by the operators $C$ and $D$ is determined entirely by
the temporal coordinate $x^0$, we identify this field as the Liouville
field $\varphi$~\cite{emn}.  Marginality of the deformation is then
restored by taking the limit $\epsilon \to 0^+$. However, in view of
(\ref{Galileanboost}), the world-sheet scale $\epsilon^{-2}$ itself is
identified with the target time~\cite{emnd,emn}, and hence one is forced
to dress the impulse operator by the Liouville/time field to restore
conformal invariance for any finite time. It is this procedure, as
explained in the relevant literature~\cite{emnd,emn} that leads to
gravitational back-reaction effects, and to an effective target-space
metric of the form (\ref{metricrec}).

The basic technical steps of this analysis may be summarised as follows.
(i) One writes the recoil boundary operator as a bulk operator in terms of
a total world-sheet derivative $\partial$, and then dresses it with a
Liouville field $\phi$. (ii) Since the $D$ operator is the leading
contribution as $\epsilon \to 0^+$, the dressed operator becomes in this
limit $\int _{\partial \Sigma} e^{\alpha \phi} \partial_\beta (U_i x^0
\Theta (x^0;\epsilon) \partial^\beta x^i) $, where $\beta = 1,2$ is a
world-sheet index, and $\alpha = \epsilon$ is the gravitational anomalous
dimension, computed according to the rules of Liouville
dressing~\cite{emnd,emn}.  (iii) One integrates this formula partially, by
applying the Stokes theorem on the world sheet, and then uses the
representation $\Theta (x^0; \epsilon)  \sim e^{-\epsilon x^0}$, where
only the zero mode of $x^0$ is used~\footnote{This is justified by a more
rigorous analysis in~\cite{emnd,szabo}.}.  (iv) In this way, one obtains
boundary contributions that are conformally invariant, as well as a bulk
world-sheet contribution of the form:
\begin{equation}\label{bulk}
U_i \int _\Sigma \epsilon x^0 \partial \phi \partial x^i e^{\epsilon(\phi 
- x^0)}.
\end{equation} 
(iv) The analysis of~\cite{szabo,emnd} indicates that ${\bar U}_i \equiv
U_i/\epsilon$ is an exactly marginal (scale-invariant) $\sigma$-model
coupling, that can be identified with the physical recoil velocity,
associated with the momentum transfer during the capture process. (v)
Taking into account the identification of the zero mode of $\phi \equiv
x^0$, and recalling that one is working at large times $\epsilon^2 x^0 
\sim
1$, one readily observes that (\ref{bulk}) corresponds to a world-sheet
interaction $\propto \partial t \partial x^i$, and hence to a modification
of the target-space background metric of the form (\ref{metricrec}). 
Physically, this is due to the displacement of the D-particle defect in 
the background space-time following the collision with the propagating 
particle.

It has been shown in~\cite{emn}
that such effects yield an effective Born-Infeld action for a massless
gauge field, which is the T-dual of (\ref{recoilop}),
corresponding to a gauge potential of the form:
\begin{equation} 
A_0 = 0 \quad (\rm gauge~condition),  \quad A_i(x^0) = 
\left(\epsilon Y_i + U_i x^0\right)\Theta(x^0;\epsilon) 
\end{equation} 
The open strings on the D-particle correspond, after a T-duality 
transformation, to gauge excitations.
We can now consider in this T-dual picture the 
propagation of a real gauge field,
represented by either a closed or open string moving through
a bulk space. A D-particle defect can be viewed
as a particular background of the corresponding $\sigma$ model. 

It has been argued in~\cite{kogwheat} that it is possible to consider
composite objects in which only one end of the open string is attached to
the D-particle, the other end being free in the bulk space.  However, as
explained in~\cite{strominger}, a fundamental string cannot end on an
isolated D-particle, because the latter does not support a world-volume
charge. Such configurations are allowed only in the presence of a
neighbouring $(p = 8)$-brane, in which case the other end of the open
string is attached to the brane.  Such a configuration experiences zero
force, which makes it an appropriate vacuum of string theory. We do not
consider such configurations here.

Instead, by analogy with the capture by the defect of a closed string, as
described above, we consider the possibility that both ends of the open
string are suddenly attached to the D-particle. In the case of isolated
D-particles, this corresponds to a composite object which transforms as an
Abelian gauge excitation under the action of the $U(1)$ group. In our
model of D-particle foam, the latter may then be identified with a
different state of the photon, differing from the incident one by a phase
in its wave function. In a similar manner, the collision with the open
string leaves the defect also with a phase difference in its wave
function.

However, the capture (or emission) of a closed- or open-string particle,
propagating through the bulk, by (or from) a D-particle defect is {\it
forbidden} for strings representing electrically-charged matter particles.
This is because, during the capture stage, which lasts for a
characteristic `uncertainty time' of order the string scale, as seen in
Fig.~\ref{fig:recoil}, an intermediate composite state is produced.  
According to the discussion above, this object transforms as a gauge
field, i.e., in the adjoint representation of the gauge group, whereas a
charged particle would in general transform according to some other
representation of the group~\footnote{In the presence of supersymmetry, 
the fermionic partners of gauge fields would behave in the same way as 
the gauge bosons themselves~\cite{Volkov}.}.

In simple terms, electric charge conservation prevents the charged
particle capture by the defect.  If the propagating closed-string probe
carried electric charge, this could not be absorbed by the defect, as it
is originally electrically neutral, as are its excitations. An excited
intermediate state is allowed {\it only} for incident photons, not for any
other Standard Model particles~\footnote{It might also be possible for
some particles beyond the Standard Model, such as right-handed (sterile)  
neutrinos, of other particles which are neutral under the Standard Model
gauge group.}. This argument certainly excludes electrons and other
elementary charged probes from having modified dispersion relations in the
model of~\cite{emn}. The discussion of composite particles such as protons
is more complicated but, to the extent that they contain uncharged
constituents such as gluons, they might also be susceptible to the effects
of the Liouville foam. We present in the next subsection a model of
space-time foam in which gluon propagation might indeed be affected.

This example of the possible breakdown of the equivalence principle in the
Liouville model of quantum-gravitational space-time foam~\cite{emn} is
just one example how the Crab Nebula synchrotron radiation constraint \cite{jlm,ems} may be evaded.

\subsection{D-Brane Model for Space-Time Foam} 

\begin{figure} [ht]
\begin{center}
\epsfig{file=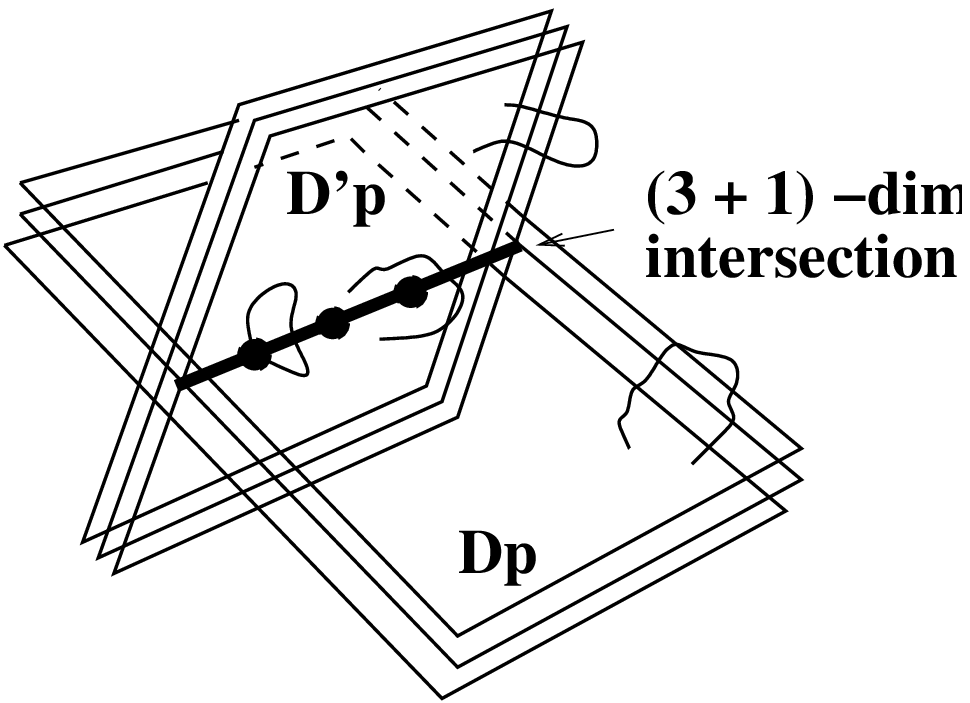,width=0.4\textwidth} \hfill 
\epsfig{file=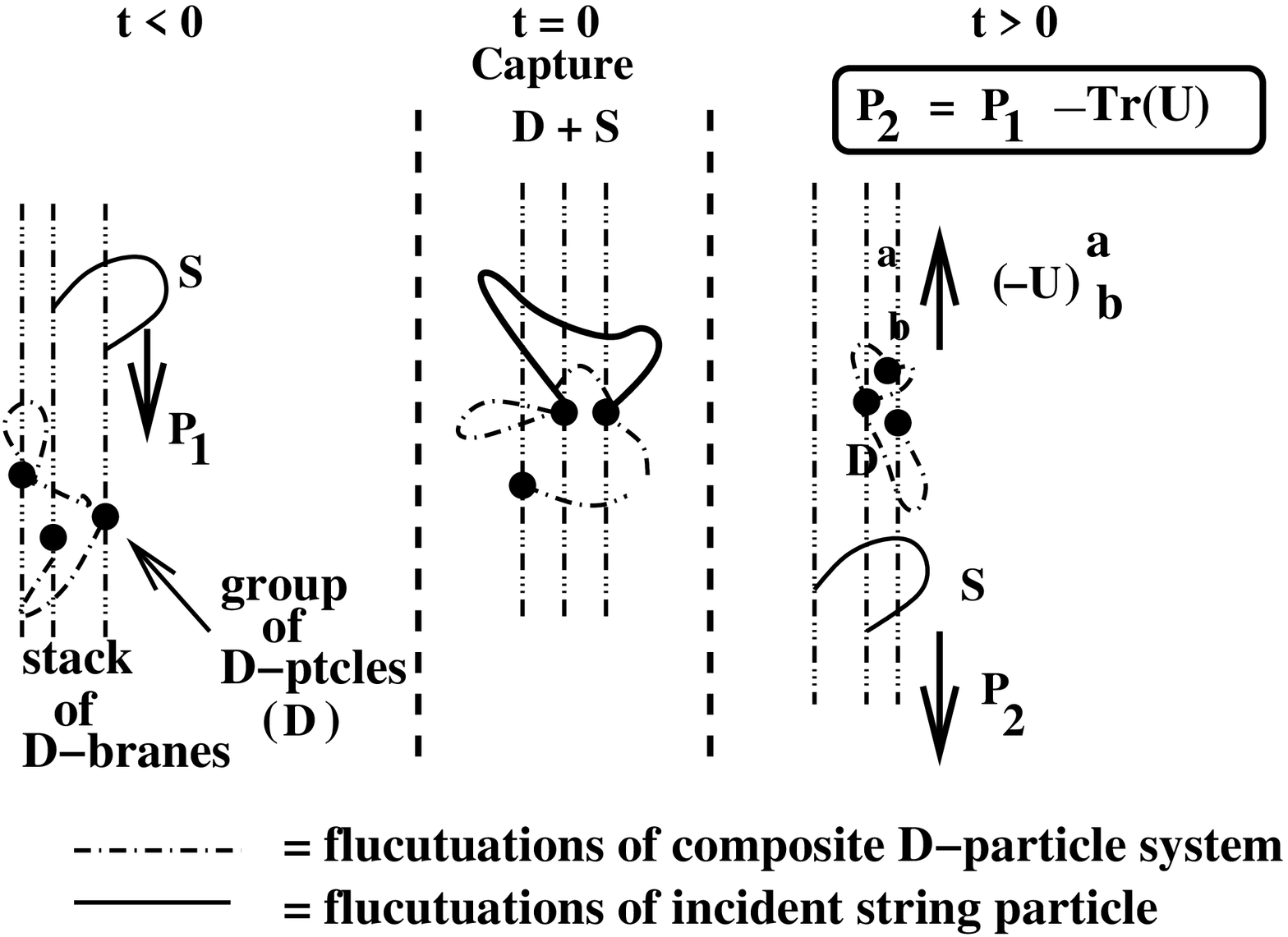,width=0.4\textwidth} 
\end{center}
\caption{\it A model of space-time foam with recoil and back-reaction which preserves a
non-Abelian gauge group. In the left drawing, the D-particles (dark blobs)  
can puncture the (3+1)-dimensional intersection of stacks of D-branes, and
the foam effect is caused by their interactions with open-string
states localised on the intersection. The right drawing shows a detailed
view of these interactions on the (3+1)-dimensional stack of intersecting
D-branes. A string mode with non-trivial non-Abelian ${\cal G}$ quantum
numbers is represented as an open-string excitation of this stack of
D-branes. Open strings can also stretch between the coincident D-particles
(as shown by the dash-dotted lines between the dark blobs), describing
excitations of groups of defects.  The recoil and back-reaction dynamics 
of a
group of $N$ interacting D-particles is described by massless gauge
excitations of non-Abelian $U(N)$ Born-Infeld type. For
virtual D-particle excitations, the group quantum numbers of the vacuum
must be preserved, so ${\cal G} \simeq U(M)$ locally, imposing selection
rules on the possible types of interaction with the foam.}
\label{non-abelian}
\end{figure}

In the modern context of the D-brane approach to gauge theories,
gauge-field excitations are viewed, not as closed-string excitations
propagating through bulk space-time, but rather as open strings with their
ends attached to a stack of $N$ Dp-branes. If these branes are parallel,
one finds excitations transforming in the adjoint representation of $U(N)$
(for oriented strings)~\cite{bound}.  If there is a stack of $N_1$
parallel Dp-(super)branes interesecting at an angle $\alpha$ with another
stack of $N_2$ parallel Dp-branes, the gauge group is $U(N_1)\otimes
U(N_2)$, and in such cases it is possible to define chiral fermions that
are localised at the interesection~\cite{angle}, which are open-string
excitations transforming under the bi-fundamental representation of the
product gauge group.  In such a way, by arranging appropriately various
stacks of higher-dimensional D-branes one may construct a representation
of the Standard Model at such a (3+1)-dimensional intersection. In this 
subsection, we re-examine the propagation of different Standard Model 
particles in such a D-brane construction.

In our approach to D-particle space-time foam we consider a situation in
which the intersection of such stacks of D-branes is punctured by
D-particle defects, as seen in the left drawing in Fig.~\ref{non-abelian}.
In such a case, the open strings localised on the brane may be captured by
the defect, as illustrated in part (b) of Fig.~\ref{fig:recoil}. The
important issue for us is: which particle species may be captured? As we
have mentioned above, in this $U(N_1)\otimes U(N_2)$ gauge theory living
on the intersection, the chiral fermions in the bi-fundamental
representation of the group and the gauge bosons in the adjoint
representation manifest themselves very differently, and it is not obvious
that any recoil/back-reaction effects should be universal.

In the case of string excitations with non-trivial quantum numbers under a
{\it non-Abelian gauge group}, one is forced to consider more complicated
foam configurations, involving groups of D-particle defects, interacting
via the exchange of open strings stretching between them. The
recoil/capture stage in such a model with non-trivial D-brane
configurations has been studied in~\cite{szabo}, where it was shown to be
described by a non-Abelian gauge theory of Born-Infeld form.

To understand physically the situation, we recall that 
the central feature of D-brane dynamics is the observation~\cite{bound} 
that the low-energy effective field theory for a system of $N$
D-branes is ten-dimensional maximally supersymmetric $U(N)$ Yang-Mills theory
dimensionally reduced to the world-volume of the D-branes. For the case of
D-particles, the Yang-Mills potential is
\beq
V_{D0}[Y]=\frac{{\cal T}^2}{4g_s}\,\sum_{i,j=1}^9\tr\left[Y^i,Y^j\right]^2
\label{D0potential}\eeq
where ${\cal T}=1/2\pi\alpha'$ is the elementary string tension, $\alpha'$
is the string Regge slope whose square root is the intrinsic string length
$\ell_s$, and $g_s$ is the (dimensionless) string coupling constant. The fields
$Y^i(t)$ are $N\times N$ Hermitian matrices in the adjoint representation and
the trace is taken in the fundamental representation of the gauge group $U(N)$.
In the free-string limit $g_s \to 0$, the field theory involving the 
potential (\ref{D0potential}) localizes onto those matrix configurations 
satisfying
\beq
\left[Y^i,Y^j\right]=0~~~~~~,~~~~~~i,j=1,\dots,9,
\label{classmatrix}
\eeq
and the D-brane coordinate fields can be diagonalized simultaneously by a
suitable gauge transformation. The corresponding eigenvalues $y_a^i$,
$a=1,\dots,N$ of the $Y^i$ then represent the collective transverse
coordinates of the $N$ D-branes. In this limit, the parallel D-branes are
very far apart from each other and massless vector states emerge only when
fundamental strings start and end on the same D-brane. The gauge group is
then $U(1)^N$. Since the energy of a string which stretches between two
D-branes is
\beq
M\propto{\cal T}\,|y_a-y_b|,
\label{stringenergy}
\eeq
more massless vector states emerge when the branes very close to each
other. The collection of all massless states corresponding to an
elementary string starting and ending on either the same or a different
D-brane forms a $U(N)$ multiplet. The off-diagonal components of the $Y^i$
and the remnant gauge fields describe the dynamics of the short open
strings interacting with the branes, with Dirichlet boundary conditions.
One obtains in this way an effective low-energy field 
theory of massless non-Abelian gauge excitations. 

We now consider the interaction of such states with groups of $N'$
D-particle defects that puncture the stack of $N$ parallel D-branes, as
appropriate for a non-Abelian extension of the D-particle foam model
discussed in the previous subsection: see Fig.~\ref{non-abelian}. The
analysis of~\cite{szabo} has demonstrated explicitly that the resulting
dynamics of the massless excitations present during the recoil/capture
stage is {\it also} described by a Born-Infeld lagrangian with $U(N')$
group.

It is useful for our purposes to repeat briefly the technical details of
the construction, so as to clarify the underlying physics. To describe the
moduli-space dynamics of a multi-D-brane system, we use the
construction described in~\cite{richfed}, which leads to the clearest
physical interpretation. In this picture, the assembly of D-branes,
including all their elementary string interactions, is regarded as a
composite `fat brane' which couples to a single fundamental string with a
matrix-valued coupling in a T-dual picture employing Neumann boundary
conditions~\footnote{For subtleties in applying the T-dual picture,
see~\cite{brecher,dornlast}.  We note here that one
assumes~\cite{richfed,szabo} that the Neumann picture is the fundamental
picture to describe the propagation of strings in fat-brane backgrounds.
The Dirichlet picture is then derived by applying T-duality as a canonical
functional integral transformation.}. The resulting effective theory is
described by a $\sigma$ model on a world-sheet with the `effective'
topology of a disc, propagating in the background of a non-abelian $U(N)$
Chan-Paton gauge field.

The relevant $U(N)$-invariant matrix $\sigma$-model action is 
\bea
S_N[X;A]&=&\frac1{4\pi\alpha'}\int_{\Sigma\{z_{ab}\}}d^2z~\tr~\eta_{\mu\nu}
\partial X^\mu\bar\partial
X^\nu-\frac1{2\pi\alpha'}\oint_{\partial\Sigma\{z_{ab}\}}
\tr~Y_i\left(x^0(s)\right)~dX^i(s)\nn\\&
&~~~~~~~~~~~~~~~~~~~~+\oint_{\partial\Sigma\{z_{ab}\}}\tr~
A^0\left(x^0(s)\right)~dX^0(s),
\label{Matrixaction}
\eea
where $\eta_{\mu\nu}$ is a (critical) flat 9+1-dimensional space-time
metric. The worldsheet fields $X$, $Y$ and $A$ are $N\times N$ Hermitian
matrices transforming in the adjoint representation of
$U(N)$~\footnote{We consider only the case of oriented open strings. For
unoriented open strings, the global symmetry group $U(N)$ is replaced by
$O(N)$ or $USp(N/2)$.}. The traces in (\ref{Matrixaction}) are taken in 
the fundamental
representation. The surface $\Sigma\{z_{ab}\}$ is a sphere with a set of
marked points $z_{ab}$, $1\leq a,b\leq N$, on it. For each $a=b$ it has
the topology of a disc $\Sigma$, while for each pair $a\neq b$ it has the
topology of an annulus. The variable $s\in [0,1]$ parametrizes the circle
$\partial\Sigma$. It was shown in~\cite{richfed} that the action
\eqn{Matrixaction} describes an assembly of $N$ parallel D-branes with
fundamental oriented open strings stretching between each pair of them.
The diagonal component $Y_{aa}$ of the matrix field $Y$ parametrizes the
Dirichlet boundary condition on D-brane $a$, while the off-diagonal
component $Y_{ab}=Y_{ba}^*$ represents the Dirichlet boundary condition
for the fundamental oriented open string whose endpoints attach to
D-branes $a$ and $b$. The matrix field $A^0$ parametrizes the usual
Neumann boundary conditions in the temporal direction of the target space.

The action \eqn{Matrixaction} is
written in terms of Neumann boundary conditions on the configuration fields,
which is the correct description of the dynamics of the D-branes in
this picture, but it is straightforward to apply a functional T-duality
transformation on the fields of \eqn{Matrixaction} to express it in the usual,
equivalent Dirichlet parametrization~\cite{richfed}. The configuration
\beq
A^\mu=\left(A^0,-\mbox{$\frac1{2\pi\alpha'}$}\,Y^i\right)
\label{gaugefield}
\eeq
can be interpreted as a $U(N)$ isospin gauge field dimensionally reduced
to the world-line of the D-brane \cite{polchinski,kss}.

As discussed in~\cite{szabo,richfed}, quantum fluctuations of the
couplings $Y_i^{ab}$ provide infinitesimal separations between the $N$
constituent D-branes proportional to the string coupling $g_s$, and also
allow for the endpoints of the fundamental strings to fluctuate in
space-time. One must then integrate out all the fluctuations among the
fat-brane constituents, i.e., over all of the marked points of
$\Sigma\{z_{ab}\}$.  This necessarily makes the action non-local. By
$U(N)$ gauge invariance, the resulting $\sigma$-model partition function
then becomes the expectation value, in a free (scalar) $\sigma$ model, of
the path-ordered $U(N)$ Wilson-loop operator $W[\partial\Sigma;A]$ along
the boundary of the world-sheet disc $\Sigma$,
\bea
Z_N[A]&\equiv&\sum_{\rm
genera}~\int[dX]~\int_\Sigma\prod_{a,b=1}^Nd^2z_{ab}~\e^{-S_N[X;A]}
\nonumber\\&\simeq&\left\langle W[\partial\Sigma;A]\right\rangle_0~\equiv~\int
Dx~\e^{-N^2S_0[x]}~\tr~P\exp\left(ig_s\oint_{\partial\Sigma}A_\mu(x^0(s))
{}~dx^\mu(s)\right),\nn\\& &~~~~
\label{partfatbrane}
\eea
where $dX$ is the normalized invariant Haar measure for integration on the Lie
algebra of $N\times N$ Hermitian matrices and
\beq
S_0[x]=\frac1{4\pi\alpha'}\int_\Sigma d^2z~\eta_{\mu\nu}\partial
x^\mu\bar\partial
x^\nu
\label{freesigma}
\eeq
is the free $\sigma$-model action for the fundamental string. The path
integral measure $Dx$ is normalized so that $\langle1\rangle_0=1$. The
partition function \eqn{partfatbrane} describes the dynamics of a fat
brane, which is depicted by the system of parallel dashed-triple-dotted
lines in Fig.~\ref{non-abelian}.

The low-energy effective action for the D-brane configuration is now obtained
by integrating out the fundamental string configurations $x$ in
\eqn{partfatbrane}. To lowest order in the gauge-covariant derivative
expansion, the result is given by a non-Abelian Born-Infeld effective 
action $Z_N[A]\simeq\e^{-N^2\Gamma_{\rm NBI}[A]}$, where
\beq
\Gamma_{\rm NBI}[A]=\frac{c_0}{\sqrt{2\pi\alpha'}\,g_s}\int dt~\tr\left({\rm
Sym}+i\zeta\,{\rm
Asym}\right)\left(\det_{\mu,\nu}\left[\eta_{\mu\nu}I_N+2\pi\alpha'
g_s^2F_{\mu\nu}\right]\right)^{1/2}
\label{nbiaction}\eeq
is the non-abelian Born-Infeld action for the dimensionally-reduced gauge
field $A_\mu$. Here $c_0$ is a numerical constant and $t=x^0(s=0)$ is the
world-sheet zero mode of the temporal embedding field, $I_N$ is the
$N\times N$ identity matrix, Sym denotes the symmetrized matrix product
\beq
{\rm Sym}(M_1,\dots,M_n)=\frac1{n!}\sum_{\pi\in S_n}M_{\pi_1}\cdots 
M_{\pi_n},
\label{Symdef}
\eeq
and Asym is the antisymmetrized matrix product
\beq
{\rm Asym}(M_1,\dots,M_n)=\frac1{n!}\sum_{\pi\in S_n}({\rm
sgn}\,\pi)M_{\pi_1}\cdots M_{\pi_n}.
\label{Asymdef}
\eeq
The symmetric and antisymmetric products on functions $f(M_1,\dots,M_n)$
of $n$ matrices $M_k$ are defined in~\cite{szabo}.  The symmetrization and
antisymmetrization operations have the effect of removing the ambiguity in
the definition of the space-time determinant in \eqn{nbiaction} for
matrices with non-commuting entries.

The components of the field strength tensor in \eqn{nbiaction} are given by
\beq
2\pi\alpha'F_{0i}=\mbox{$\frac
d{dt}$}\,Y_i-ig_s[A_0,Y_i]~~~~~~,~~~~~~(2\pi\alpha')^2F_{ij}=g_s[Y_i,Y_j]
\label{fieldstrength}
\eeq
and the constant $\zeta\in\real$ is left arbitrary so that it interpolates
among proposals for the true trace structure inherent in the
non-Abelian generalization of the Born-Infeld action. The case $\zeta=0$
corresponds to the original proposal in~\cite{tseytlin}, while the trace
structure with $\zeta=1$ was suggested, in a different context,
in~\cite{argnappi}.  The two-loop world-sheet $\beta$ function for the
model \eqn{partfatbrane} was calculated in~\cite{brecher} to be
\beq
\beta_i^{ab}\equiv\frac{\partial
Y_i^{ab}}{\partial\log\Lambda}=-(2\pi\alpha'g_s)^2\left(D^\mu F_{\mu
i}\right)^{ab}+2(2\pi\alpha'g_s)^3\left(D^\mu\left[F_{\mu\nu},F^\nu_{~i}
\right]\right)^{ab}+{\cal O}\left((\alpha'g_s)^4\right),
\label{betamatrix2loop}
\eeq
where
\beq
D_0=\mbox{$\frac d{dt}$}-ig_s\left[A_0,\cdot\,\right]~~~~~~,~~~~~~
D_i=\mbox{$\frac{ig_s}{2\pi\alpha'}$}\left[Y_i,\cdot\,\right]
\label{gcderivsred}
\eeq
are the components of the dimensionally-reduced gauge-covariant 
derivative. It
is readily seen that \eqn{betamatrix2loop} coincides with the variation of the
action \eqn{nbiaction} with $\zeta=1$ up to the order indicated in
\eqn{betamatrix2loop}, so that the world-sheet RGEs $\beta_i^{ab}=0$
coincide with the equations of motion of the D-branes. The
first term in \eqn{betamatrix2loop} yields the (reduced) Yang-Mills equations
of motion, while the second term represents the first-order stringy 
correction to the Yang-Mills dynamics. 

If we now replace the space-time 
coordinates $Y_i$ and velocity $U_i$ used in the previous subsection by 
matrix quantities:
\beq
Y_i(x^0)^{ab}=Y_i^{ab}+U_i^{ab}\,x^0,
\label{matrixboost}
\eeq
the discussion of D-particle recoil in the previous subsection may be
generalized to the D-brane model illustrated in Fig.~\ref{non-abelian},
where the intersection is viewed as a foam model of virtual D-branes
emerging from the vacuum. Because of this assumption, the D-particle
defects are not real, but appear as virtual excitations of the foam, and
as such are bound to carry the same quantum numbers as the vacuum.

However,
a model where
the bulk ten-dimensional space, in which the D-branes are embedded,
is characterised by a given density
of such D-particle defects, occasionally crossing
the intersection, is also a viable model
for space time foam. To an observer on the intersection,
the crossings of the D-particle defects will appear as
`virtual' D-particle fluctuations in the vacuum.
Formally, the (T-dual) version of the recoil operator
describes an effective non-Abelian gauge potential
of the form~\cite{szabo}:
\begin{equation}
A_0^{ab} =0~; \quad A_i^{ab} = \left(\epsilon Y_i^{ab}+U_i^{ab}\,x^0
\right)
\Theta (x^0;\epsilon)
\label{NArecoilop}
\end{equation}
where $a, b$ are a colour indices. As shown in~\cite{szabo},
its dynamics is that of a non-Abelian Born-Infeld
described by the lagrangian (\ref{nbiaction}).

Thus, in analogy with the Abelian case of an isolated D-particle,
the recoil and capture of string matter by groups of $N'$ such
defects results in gauge excitations transforming in the adjoint
representation of the $U(N')$ group. For real defects, this group might
not be connected with the original gauge group of the string excitations
localised on the intersection. However, given that in our model of foam
the D-particles are viewed as virtual excitations of the ground state of
our system, and we insist on the possibility of re-emission of the string
matter into the (stacked) brane world after the capture stage, the two
groups must be the same. In other words, the number $N'$ of the
constituent D-particles in each excitation of the non-Abelian foam has to
be such that the resulting $U(N')$ group must be locally isomorphic to the
$U(N_1) \otimes U(N_2) \otimes \dots $ group coming from the the stacks of
the intersecting D-branes. This means that the gauge boson configurations
(\ref{recoilop}) that emerge from the recoil/capture stage described in
the previous subsection must here be thought of as parts of a
superposition with the gauge boson excitations arising from open strings
on the intersecting stack of D-branes, realizing the same gauge group.

It follows from these considerations that the above procedure describes
only the interaction of {\it massless} gauge particles (and their
superpartners), such as Standard Model photons and gluons, with the
space-time foam. There is no possibility of a non-trivial interaction of
electrically-charged massive particles, such as electrons, with the
D-particle defects. As in the Abelian case, this has to do with the fact
that such charged probes transform according to bi-fundamental
representations of the $U(N_1)\otimes U(N_2)\dots $ group, whilst the
composite open-string/D-particle excitations transform in the adjoint
representation, and hence behave like gauge particles. The D-particle foam
medium is transparent to such elementary matter particles, because of the
conservation of electric charge, which cannot be broken by the
gravitational space-time foam. We stress that this picture of space-time
foam is not covered by critical string theory arguments, because recoil
operators (\ref{NArecoilop}) are ignored in such treatments. In our
approach, it is the presence of such non-conformal, relevant operators
that leads to the non-trivial non-equilibrium space-time foam effects for
photons that we have examined here.

The above possible effects of non-Abelian gauge interactions of D-particle
groups may be suppressed significantly for composite particles containing
gluons, such as protons.  As argued in~\cite{ng,emngzk}, propagation
effects and energy uncertainties in collision processes due to the
Liouville foam interactions of ultra-high-energy protons, might be
responsible for the possible persistence of ultra-high-energy cosmic
rays~\cite{piran} beyond the expected GZK cutoff in the spectrum. However,
as observed in~\cite{emngzk}, the parameter ${\cal M}$ needed to modify
the proton dispersion relation would be of order $10^{-16}-10^{-17}$
smaller than the Planck mass, much smaller than the corresponding
parameter (of order one in Planck units) that would be sufficient to
remove the expected analogous cutoff for gamma rays above 20
TeV~\cite{PM}\footnote{Some doubt~\cite{PM,steckerblazars,kran} has been
expressed on whether a TeV-infrared background crisis exists. This is due to
ambiguities in measurements of extragalactic light~\cite{kran}, as well as
theoretical uncertainties in the intrinsic model for generation of TeV photons
in Active Galactic Nuclei. In our view, observations of photons in the
100-200~GeV energy range, as well as TeV sources at different redshifts, are
needed in order to constraint seriously space time foam models with modified
photon dispersion relations.}. Such a difference might well be explained by the
fact that
only the gluons in the proton feel the effects of the D-particle foam.
Since the characteristic scale of a QCD bound state suach as the proton is
of order $\Lambda_{QCD} \sim 200$ GeV, there could be an extra suprression
factor $\Lambda_{QCD}/M_P \sim 10^{-19}$ multiplying the coefficient of
$E/{\cal M}$ in the modification of the proton dispersion relations.
However, more work is required before firm conclusions can be reached
regarding such issues in the context of the model of~\cite{emn}.
Nevertheless, the above analysis makes it clear that details in the
dynamics of the interactions between foam and matter can make a crucial
difference from naive generic phenomenological analyses, which could be
misleading.

\section{Discussion}

In light of the analysis in the present work, it may still be that the
study of the arrival times of photons in gamma-ray bursts
(GRBs)~\cite{gray} is the best experimental test, to date, for probing
possible quantum-gravitational effects on particle propagation, through
their high sensitivity to any possible refractive index for photons
propagating {\it in vacuo}.

We have pointed out previously that ultra-high-energy neutrino emission
from GRBs would also be very sensitive to any analogous effects on
neutrino propagation~\cite{Volkov}. However, in the type of
intersecting-brane model studied here, neutrinos are considered as chiral
fermions that transform in bi-fundamental representations of the
weak-interaction gauge group $SU(2)_L \otimes SU(2)_R$. As such, they do
not
interact, at least at tree level, with D-particles. However, this
conclusion, stemming from the analysis of~\cite{szabo} described above,
applies only to the case in which the open string is attached with both
ends on the D-particle, and hence the open strings have double Dirichlet
(DD) boundary conditions.

\begin{figure} [ht]
\begin{center}
\epsfig{file=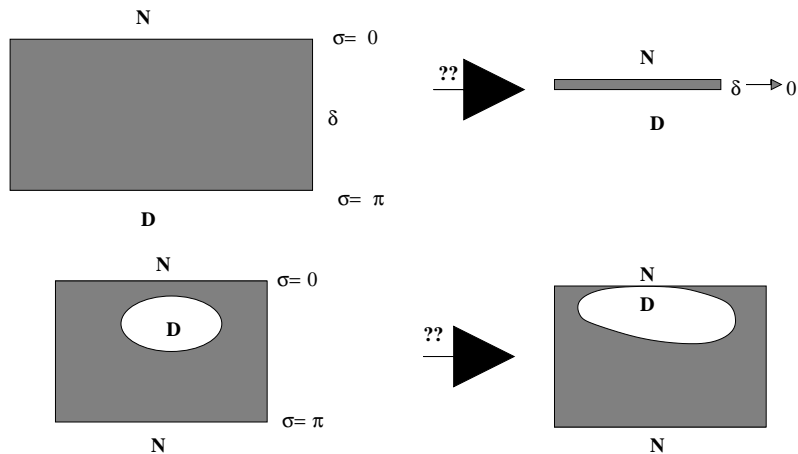,width=0.7\textwidth}
\end{center}
\caption{{\it Top drawing: if ND open strings were consistent conformal
invariant backgrounds, then the pinching of world-sheet surfaces
should be allowed. However, this would require the superimposition of
different types of boundary conditions, which is impossible.
This argument also rules out the appearance of
mixed boundary conditions for open strings exchanged in
string loops, as seen in the bottom drawing.
Quantum string higher-genus corrections to ND strings
are inadmissible, as pinched genera in the loop would also require
a superimposition of two different types of boundary conditions 
in the limit where the world-sheet strip width vanishes.}}
\label{pinched}
\end{figure}

This raises the question whether, in principle, one may consider the
recoil of strings with one end attached by a Dirichlet boundary condition
to the D-particle and the other attaches freely with a Neumann boundary
condition to intersecting D-branes, as shown in Fig.~\ref{non-abelian}. If
such ND strings are consistent string configurations, as argued
in~\cite{kogwheat}, the earlier discussion about the lack of Lorentz
violation for charged particles such as electrons would be unaffected,
though one could apply the ND formalism to neutrinos. However, in our view
there is a issue in defining such configurations, associated with the
issue of conformal invariance for pinched higher-genus configurations, as
illustrated in Fig.~\ref{pinched}. If such open ND strings existed, there
is a question whether they could be consistent configurations, from a
target space-time point of view. Because of the mixed ND boundary
conditions, when one pinches a higher-genus world-sheet graph, the pinch
would contain a world-sheet strip with infinite length and vanishing
width, on the ends of which one would impose two different boundary
conditions. This would require the introduction of a non-local phase-space
operator, called a `zipper' in~\cite{kogwheat}. It remains to be
established whether this is a consistent local operator in conformal field
theory.

Another important issue concerns higher-order self-energy graphs. The
naive (local field theory) point of view would be that the quantum
fluctuations of open strings interacting with D-particles are described by
annulus or higher-topology world-sheet graphs, involving the exchange of
all string states around the loop. In principle, some such states might
interact with the D-particle. However the same argument based on conformal
invariance that was given above for pinched tree-level world sheets also
applies here. This can be seen, in particular, by considering the lower
part of Fig.~\ref{pinched}, featuring a chiral fermion with Neumann
boundary conditions and a string loop with Dirichlet boundary conditions,
representing quantum corrections in which a D-particle interacts with the
virtual string state. We conclude that the non-interaction of chiral
fermions with the D-particle foam is an exact result, valid to all orders
in the string genus expansion.

However, as discussed above, gauge fields may in general interact with the
space-time foam, leading to violations of Lorentz invariance that are
non-universal and generating violations of the equivalence principle at
high energies that may be probed by astrophysical observations.

\section*{Acknowledgements}

We would like to thank H.~Hofer for his interest and support. The work of
N.E.M. is partially supported by the European Union through contract
HPRN-CT-2000-00152. The work of D.V.N. is supported by D.O.E. grant
DE-FG03-95-ER-40917.

\end{document}